\documentclass[preprint,8pt]{elsarticle}

\usepackage{amssymb,amsmath,amsfonts}
\usepackage{graphicx}
\usepackage{lineno}
\usepackage{caption}
\usepackage{multicol}
\usepackage{multirow}
\usepackage{adjustbox}
\usepackage{floatrow}
\usepackage{color}
\usepackage{bm}
\usepackage{ulem}
\usepackage{enumitem}
\newlist{subquestion}{enumerate}{1}
\setlist[subquestion,1]{label=(\alph*)}
\usepackage{float}
\usepackage{geometry}

\usepackage{placeins}

\newcommand{\ie}{\textit{i.e., }}

\usepackage{hyperref}
\usepackage{breqn}

\journal{Acta Materialia}

\begin{document}

\begin{frontmatter}
\title{How glass breaks -- Damage explains the difference between surface and fracture energies in amorphous silica}

 \author[lamcos]{Gergely Moln\'{a}r\corref{cor1}}
 \ead{gergely.molnar@insa-lyon.fr}

 \author[espci]{Etienne Barthel}

\cortext[cor1]{Corresponding author}

\address[lamcos]{CNRS, INSA Lyon, LaMCoS, UMR5259, 69621 Villeurbanne, France}
\address[espci]{Soft Matter Sciences and Engineering, CNRS, ESPCI Paris, PSL University, Sorbonne Université, 75005 Paris, France}

\begin{abstract}
      The relationship between free surface energy and fracture energy in amorphous silica is investigated using multiscale simulations. We combine the homogenization of an atomic scale fracture model with a phase-field approach to track and quantify the different energy contributions. Free surface energy, localized as potential energy at the surface, is clearly separated from damage diffusion extending over a 16-23~\r{A} range around the crack path while the plastic contribution is found negligible. Analysis of the local structure reveals that diffuse damage is primarily associated with changes in the ring structure, whereas variations in free surface energy correlate with changes in the coordination number of silicon atoms.
      These findings provide new insight into brittle fracture mechanisms in amorphous silicates and are consistent with experimental measurements of silica toughness.
\end{abstract}

\begin{keyword}
Free surface energy \sep Fracture toughness \sep Phase-field \sep Molecular dynamics \sep Internal length scale \end{keyword}

\end{frontmatter}

\section{Introduction}
\label{sec:introduction}

Surface free energy quantifies the excess energy associated with unbalanced interatomic bonds at a material's surface. In solids, the bulk state is always energetically more favorable, therefore creating a surface requires an energy input. This quantity is an essential parameter in surface science, as it governs how material surfaces directly influence phenomena such as wetting, adhesion, coating, friction, and corrosion.

On the other hand, Griffith \cite{Griffith1921} postulated a closely related concept that forms the basis of fracture mechanics and is essential for estimating fracture resistance \cite{Mueller2015}. He stated that a free surface can be created and a crack can propagate when the elastic energy released exceeds a material-specific critical value. Essentially, both free surface energy and fracture surface energy describe the same underlying physical phenomenon and in an ideal brittle material, the fracture energy is expected to equal twice the surface energy.

However, in experiments, fracture surface energy is almost always significantly larger. It is the case for polymers \cite{Tokuda2015,Smith2022} and crystalline metals \cite{Bikerman1978} where the fracture energy exceeds the free surface energy by several orders of magnitude due to plastic dissipation around the crack tip, as well as for bulk metallic glasses \cite{Xu2010,Yiu2020}. One of the rare exceptions is silicon crystals \cite{Gilman1960,Perez2000,Eaglesham1993}, where the difference falls within the precision limits of the methods used, and which can be assumed to break by individual bond rupture at an atomically sharp tip as expected from a brittle material~\cite{Gumbsch1995},

Intriguingly, for silicate glasses, the archetypes of brittle materials, it has long been known that the measured fracture energy exceeds the surface free energy by a factor of ca. 5~\cite{Wiederhorn69,Rhee1977,Quinn2017}. The difference has been tentatively ascribed to plastic dissipation~\cite{marsh1964plastic,Wiederhorn69} but as it is difficult to experimentally evidence plasticity at this scale in an amorphous material, let alone quantify it (and also because it openly conflicts with the accepted notion of brittleness), the claim has remained controversial~\cite{lawn1980atomically} to these days: signs of plasticity at silica crack tips have been discounted~\cite{Guin2004} almost as soon as found~\cite{Celarie2003}. Worse still, standard models for plastic dissipation in cracks~\cite{Wei99} are inapplicable to silicates because their plasticity is essentially devoid of hardening~\cite{Kermouche16}.

Faced with this conundrum, we focused on silica glass to investigate the rupture of brittle amorphous silicates at the crossover between the atomic and the continuum scales. Previous atomic-scale simulations have recognized the existence of an inelastic zone around the crack tip \cite{Rountree2007,Rimsza2018} and have estimated its size; however, they did not provide a mechanical explanation for its origin, nor did they describe its structural signature.

Here, we show that we can clearly separate the two contributions to the fracture energy : 1) the energy required for surface formation—accounting for bond breaking and atomic relaxation near the surface—, and 2) the energy due to structural rearrangements extending much deeper into the material, over a distance of approximately 20 \r{A}. We also show that the latter process is best described as damage rather than plasticity or nonlinear elasticity and that the relative contributions match the experimental results. This approach offers a fully consistent framework for understanding fracture in amorphous silica, an archetype of brittle materials whose fracture mechanisms exhibit unexpected complexity.

While the macroscopic effect of crack-tip regularization is well known \cite{Irwin1958,Bazant1999,Leguillon2018,Molnar2020TAFMEC}, its spatial extent is rarely characterized due to the complexity of experimental observations and the fact that existing studies are mostly focused on highly inhomogeneous materials \cite{Cedolin1983,Brooks2013}. As a consequence, this paper is among the first to not only demonstrate the effect of this diffuse damage, but also to identify its spatial extent and provide a structural explanation for its existence. In doing so, it provides one of the first direct demonstrations of the diffuse damage concept employed in recently popular phase-field fracture techniques \cite{Bourdin2000,Ruffini2015}.

\section{Methods}
\label{sec:methods}

To explore the fracture properties of amorphous silica, the atomistic simulations employed the BKS potential \cite{Yuan2001} to model atomic interactions and Wolf's truncation method to handle Coulombic forces \cite{Carre2007}. The BKS potential was shown to qualitatively but consistently describe the elastic and plastic mechanical properties of these systems \cite{Molnar2016AM,Molnar2017MM}. Amorphous glass samples were generated by randomly distributing atoms within a periodic box. Molecular dynamics simulations were then performed to equilibrate, quench, and test the samples. Athermal (quasi-static) deformations, pertinent for systems below the glass transition temperature, were applied through iterative energy minimization using the conjugate gradient method \cite{Molnar2016AM}. 

Two types of simulations were conducted: (i) fracture propagation analysis; and (ii) free surface energy calculations.

\subsection{Sample generation}

The amorphous glass samples were prepared through a random sequential placement of atoms within a periodic simulation box. Following this, molecular dynamics simulations using the LAMMPS software~\cite{LAMMPS} were conducted to equilibrate, quench, and test the samples. In this study, we examined two sample sizes: a cubic box with a side length of 100~\r{A}, and a rectangular cuboid with dimensions of $400 \times 300 \times 100$~\r{A}$^3$. After equilibration at 3000 K, the samples were quenched at a rate of 10 K/ps in an NPT ensemble to a final temperature of 10$^{-5}$ K. The samples were then deformed in an athermal manner. The sample generation methodology and the potential function employed are detailed in Ref.~\cite{Molnar2016JNCS}.

\subsection{Mechanical deformation}

Fracture simulations were carried out on a 3D sample with dimensions $L_x \times L_y\times L_z=400 \times 300 \times 100$ \r{A}$^3$, containing 800k atoms, which was found to be large enough to capture initiation and  propagation adequately. The initial defect, a rounded incision extends along the $z$ direction on a face normal to the $x$ direction. There is periodicity in the $z$ direction. For the boundary conditions on the $x$ and $y$ faces, a K-field displacement was imposed, simulating uniaxial tension along $y$ at infinity. The crack then initiated and propagated along the $x$ direction as the loading parameter ($K_I$) was increased. 

\begin{figure}
	\centering
	\includegraphics[width=0.99\textwidth]{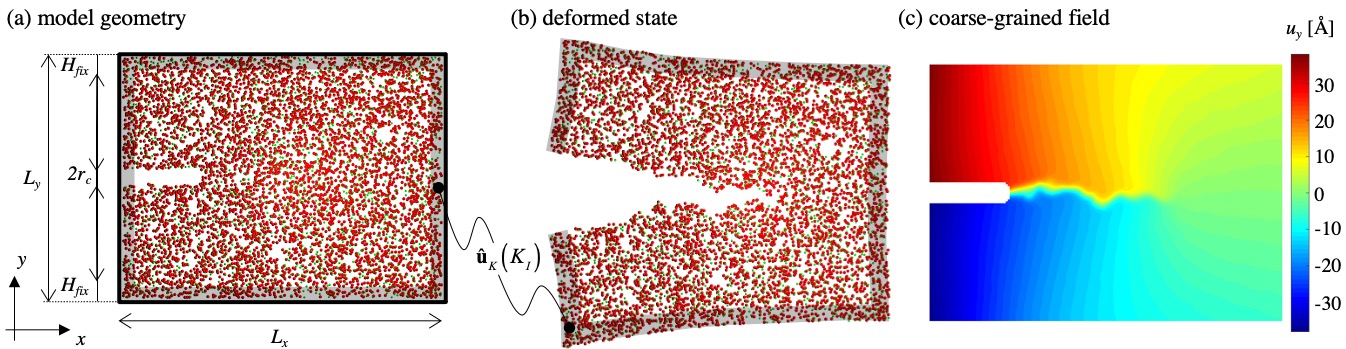}
	\caption{(a) A cut of the atomic-scale model displaying atoms of a 5~\r{A} thickness. (b) Deformed configuration with the same cut with $K_I=2.2$~MPa$\sqrt{m}$.
		(c) Coarse-grained displacement field in the Lagrangian configuration, obtained via convolution, at the same loading state as in (b).}
	\label{fig:Model}
\end{figure}

To simulate crack propagation, the periodic boundaries in the $x$ and $y$ directions were removed. An initial crack with a length of 100~\r{A} and a radius of $r_c=10$~\r{A} was introduced. The outer free boundaries, fixed at a distance of $H_{\rm{fix}} = 15$~\r{A}, were displaced iteratively over 4000 steps using the following K-field until $K_I=2.2$~MPa$\sqrt{m}$:

\begin{equation}
	\label{eq:Kfield}
	\begin{array}{l}
		{{\hat u}_x} = \frac{{{K_I}}}{{8\mu \pi }}\sqrt {2\pi r} \left[ {\left( {2\kappa  - 1} \right)\cos \frac{\theta }{2} - \cos \frac{{3\theta }}{2}} \right],\\
		{{\hat u}_y} = \frac{{{K_I}}}{{8\mu \pi }}\sqrt {2\pi r} \left[ {\left( {2\kappa  + 1} \right)\sin \frac{\theta }{2} - \sin \frac{{3\theta }}{2}} \right],
	\end{array}
\end{equation}

\noindent
where $r$ and $\theta$ are polar coordinates measured from the initial crack tip, $\mu$ is the shear modulus, $\kappa=3-4\nu$, and $\nu$ is Poisson's ratio. For the bulk sample, the elastic constants were determined as $\mu=31.7$~GPa and $\nu=0.25$. Note, that the largest displacement increment was smaller than 0.01~\r{A}, which is less than 1~\% of the characteristic inter-atomic distance ($r_{Si-O} = 1.61$~\r{A}). This ensured that the applied deformation remained independent of the step size. Note, that $H_{\rm{fix}} = 15$~\r{A} is larger than the interatomic potential cutoff. A schematic illustration of the initial sample and the deformed configuration is depicted in Fig.~\ref{fig:Model}a and b, with a cut in the $z$ direction. During the progressive advance of the crack tip, $K_I$ acts solely as a loading parameter and does not retain its intrinsic physical meaning.

\subsection{Continuum damage}

Local continuum quantities were computed using a physically based Gaussian convolution technique, commonly known as the coarse-graining method  \cite{Hardy1982,Goldhirsch2002}. This method offers the advantage of conserving both energy and mass during the homogenization procedure. With this technique, homogeneous fields for displacements, displacement gradients, strains, rotations, stresses, and potential energy were obtained for both the initial (Lagrangian) and deformed (Eulerian) configurations. As a result, the following continuum quantities were available in both Lagrangian and Eulerian configurations: displacements $u$, Hencky (logarithmic) strain tensor $\bm{\varepsilon}$, Cauchy stress tensor $\bm{\sigma}$ , potential energy density $\psi_{\rm{pot}}$ (sum of the atomic interactions) and mass density $\rho$.

To execute the convolution, the following function was employed:

\begin{equation}
	\label{CGfun}
	\phi \left( r \right) = \frac{1}{{{w^3}{\pi ^{3/2}}}}{e^{ - \frac{{{r^2}}}{{{w^2}}}}},
\end{equation}

\noindent
where $r$ is the distance between the observation point and the atom, and $w$ is the coarse-graining width. This function is normalized such that its integral in 3D equals 1. More details about the technique can be found in Refs.~\cite{Molnar2016JNCS,Molnar2017PRE}. An example of the coarse-grained displacement in the $y$ direction is illustrated in Fig.~\ref{fig:Model}c. The choice of the CG width $w$ will be discussed in detail below. 

We recall that from the simulations, the coarse-grained stress, strain, and local initial stiffness fields are available. We can calculate damage in the mechanical sense \cite{Kachanov1958} as the difference between the actual coarse-grained elastic energy ($\psi_{el}$) and the undamaged energy components obtained from the strain field:

\begin{equation}
	\label{eq:psiel}
	{\psi _{el}}\left( {{\bf{u}},d} \right) = {\left( {1 - d} \right)^2}\psi _0^ + \left( {\bm{\varepsilon} \left( {\bf{u}} \right)} \right) + \psi _0^ - \left( {\bm{\varepsilon} \left( {\bf{u}} \right)} \right),
\end{equation}

\noindent
with $d$ representing the damage, $\psi _0^ + $ and $\psi _0^ - $ representing the tensile and compressive parts of the undamaged strain energy densities. Note that we assume there is no damage in compression. We use the spectral energy decomposition \cite{Molnar2022EFM} here, because we found a comparable failure surface in crack free deformations than in atomic scale simulations \cite{Molnar2016AM}. We note, that more advanced degradation functions exist and can affect the quantitative value of the phase-field regularization length.

Thus $d$ evolves between 0 for the pristine and 1 for the fully damaged material. This variable represents the progressive loss of stiffness and load-bearing capacity of the material.

Under the assumption of linear elasticity, we can calculate the elastic strain energy density field~\footnote{Capital $\Psi$ is the global value integrated over the whole volume ($V$), while small $\psi$ refers to local energy densities.} as
\begin{equation}\label{eq:psi_eps}
	\psi _{\epsilon}~=~{\psi _{\epsilon,0}} + \int_{{\bm{\varepsilon} _0}}^{\bm{\varepsilon}} {\bm{\sigma} :d\bm{\varepsilon} }  \approx {\psi _{\epsilon,0}} + {\textstyle{1 \over 2}}\bm{\sigma} :\bm{\varepsilon} J,
\end{equation} where $J$ denotes the determinant of the deformation gradient tensor, accounting for volume changes, and $\psi_{\epsilon,0}$ is the initial strain energy density, which may be non zero, typically after quenching.

The local elastic moduli were calculated by deforming the large sample in 6 elementary ways incrementally by $\delta\epsilon = 0.01$~\% until 0.5~\% strain (3 compression, 3 shear). The local stresses were then correlated to local strains. This method, as used in Ref.~\cite{Molnar2016JNCS} to correlate local soft spots to sodium distributions, employs the same numerical parameters. Further details are available in the aforementioned paper.

Damage is calculated in the Lagrangian configuration following principles of solid mechanics, as most quantities are better defined in the initial state. Only Cauchy stresses needed to be interpolated back from the deformed state because coarse-graining atomic pairs in the initial configuration that were no longer in contact often resulted in locally negative strain energy, which is physically impossible. This interpolation involved displacing grid points of the Lagrangian configuration by their coarse-grained displacements, followed by interpolating stress values from the Eulerian grid to these displaced points.

To account for free surfaces in the Eulerian configuration that appear on the crack lips, a correction multiplier was defined based on the ratio of locally interpolated densities from the deformed configuration to those from the initial configuration: $\xi  =  {\rho _L} / \rho _E^{\text{int}} $. This adjustment compensates for the absence of material at the free surfaces in the deformed configuration, ensuring accurate calculations when parts of the coarse-grained volume are empty.

As amorphous materials lack the ordered structure of crystals, initial local stresses can be found in the quenched material. We assume that these stresses store elastic energy, which is quantified and added to the deformation calculated during loading. The initial strain field can be calculated using linear elasticity as: ${\bm{\varepsilon}_0} = {\bm{C}^{-1}}{\bm{\sigma}_0}$, where $\bm{C}$ is the local rigidity tensor and $\bm{\sigma}_0$ is the local initial stress tensor. The initial elastic strain energy density was then obtained by ${\psi_{el,0}} = \frac{1}{2} \bm{\sigma}_0 : \bm{\varepsilon}_0$. We note that this quantity was significantly smaller than the energy from the applied deformation.

\subsection{Free surface energy}

To quantify the energy required to create a free surface, we employed a standard approach from the literature \cite{Rimsza2018}. A smaller sample was cut at different positions by displacing atoms along the $y$ direction in a periodic environment, after which the periodic boundary condition in the $y$ direction was replaced with free surfaces. The energy of the sample was then minimized, and $2\gamma$ was calculated by dividing the energy difference by the area of the newly created surfaces:  $2\gamma  = {{\Delta {\Psi _{pot}}} \mathord{\left/ {\vphantom {{\Delta {\Psi _{pot}}} {\left( {{L_x}{L_z}} \right)}}} \right.
		\kern-\nulldelimiterspace} {\left( {{L_x}{L_z}} \right)}}$.

Furthermore, after energy minimization---performed in the same manner as for the mechanically deformed sample---local quantities were evaluated. Owing to the originally periodic nature of the problem, coarse-graining was carried out in the Lagrangian configuration, and continuity was assumed at both free surfaces to prevent averaging over the empty volume surrounding the sample.

\subsection{Finite element scheme}

In finite element formulations, damage is typically obtained by minimizing the total energy of the model with respect to the unknown degrees of freedom, namely the local damage variables. In the case of the phase-field approximation, the material properties---such as the critical fracture toughness ($g_c$) and the internal length scale---are assumed to be known. Given the crack-driving force, i.e., the undamaged strain energy density, the governing differential equation can be solved to obtain the local damage field.

In this section, in contrast to the standard implementation, we determine the set of material parameters that best fits the damage field obtained using coarse-grained analysis by minimizing the difference between this field and that computed using a finite element phase-field model.

To show how the discrete results can be fed into a continuum scale description, the damage field was recalculated using a Finite Element Update (FEMU) \cite{Kavanagh1971} scheme on 20 slices of the sample to capture the variation along the thickness. During this process, the fracture toughness $g_c$ and a constant damage width $l_c$ were fitted to minimize the differences between the damage values obtained from coarse-grained molecular scale simulations and those calculated through a phase-field approach~\cite{Bourdin2000}. We assumed that the undamaged strain energy density $\psi_0$ is known at the atomic scale and minimized the internal energy of the system expressed as:

\begin{equation}
	\label{eq:EngInt}
	{\Psi _{{\mathop{\rm int}} }}\left( {{\bf{u}},d} \right) = \int_V  {{\psi _{\epsilon}}\left( {{\bf{u}},d} \right)dV }  + {g_c}\Gamma \left( {d,\nabla d} \right),
\end{equation}

\noindent
where $\Gamma$ represents the crack surface. In the phase-field approach \cite{Pham2011}, it is approximated using a crack density function ($\gamma_\Gamma$):

\begin{equation}
	\label{eq:PFSurf}
	\Gamma  = \int\limits_V {{\gamma _\Gamma }\left( {d,\nabla d} \right)dV}  =  \frac{3}{{8{l_c}}}\int_V  {\left( { d  + l_c^2{{\left| {\nabla d} \right|}^2}} \right)dV }.
\end{equation}

The objective of the technique is to update the parameters of a constitutive model so that the results of the phase-field simulation, under appropriate boundary conditions, match as closely as possible the results obtained through atomic scale simulations in the sense of a given norm. We used an AT1 description for the phase-field model with a quadratic degradation function.

The FEMU method utilizes the undamaged tensile energy ($\psi_0^+$) to obtain the local phase-field damage variable. The approach involves iteratively adjusting the material properties, which are considered homogeneous in this case, to minimize the difference between the damage field obtained from the ratio of damaged to undamaged energies ($\bf{d}_{\rm{MD}}$) and the damage field from finite element calculations ($\bf{d}_{\rm{FEM}}$).

\begin{equation}
	\label{eq:FEMU}
	\bm{\Lambda}  = {\rm{arg min}}{\left[ {{{\bf{d}}_{{\rm{MD}}}} - {{\bf{d}}_{{\rm{FEM}}}}\left( \bm{\Lambda} \right)} \right]^T}\left[ {{{\bf{d}}_{{\rm{MD}}}} - {{\bf{d}}_{{\rm{FEM}}}}\left( \bm{\Lambda}  \right)} \right],
\end{equation}

\noindent
with  $\bm{\Lambda}  = \left[ {\begin{array}{*{20}{c}}{{g_c}}&{{l_c}}\end{array}} \right]$. The iteration is done by solving the following linear equation system:

\begin{equation}
	\label{eq:FEMUit}
	{\bf{M}}d\bm{\Lambda}  = {\bf{b}},
\end{equation}

\noindent
with

\begin{equation}
	\begin{array}{l}
		{\bf{M}} = {\left[ {\frac{{\partial {{\bf{d}}_{\rm{FEM}}}}}{{\partial \bm{\Lambda}  }}} \right]^T}\left[ {\frac{{\partial {{\bf{d}}_{\rm{FEM}}}}}{{\partial \bm{\Lambda}  }}} \right],\\
		{\bf{b}} = {\left[ {\frac{{\partial {{\bf{d}}_{\rm{FEM}}}}}{{\partial \bm{\Lambda}  }}} \right]^T}\left[ {{{\bf{d}}_{\rm{MD}}} - {{\bf{d}}_{\rm{FEM}}}\left( \bm{\Lambda}  \right)} \right].
	\end{array}
\end{equation}

The fracture properties were changed until the maximum change in error was smaller than $10^{-6}$. The procedure was executed for 20 equally spaced 2D slices in the $z$ direction.

In the literature~\cite{Rimsza2018}, crack length is often calculated based on an assumed position of the crack tip, either from local density or using a singular Williams series fit. These methods are not sufficiently precise to describe the moving crack. Furthermore, they cannot capture multiple crack fronts. Here the FEMU approach provides a first method to determine the crack length which overcomes these drawbacks.

As a result, we were able to obtain the crack length, the fracture toughness, and the internal length scale of the material from a single snapshot of deformation without accounting for the boundary conditions, relying solely on local quantities.

\FloatBarrier
\section{Results}

\subsection{Diffuse damage}

In the case of purely elastic deformation, the local elastic energy density field $\psi _{\epsilon}$ is equal to the undamaged strain energy $\psi _0$ calculated from the strain and the initial local elastic stiffness. In contrast, our results evidence a dramatic reduction of the local elastic energy density $\psi _{\epsilon}$ over a large area surrounding the crack faces. This result is evidence for a significant evolution of the structure of the material around the crack tip.
\begin{figure}
	\centering
	\includegraphics[width=0.5\textwidth]{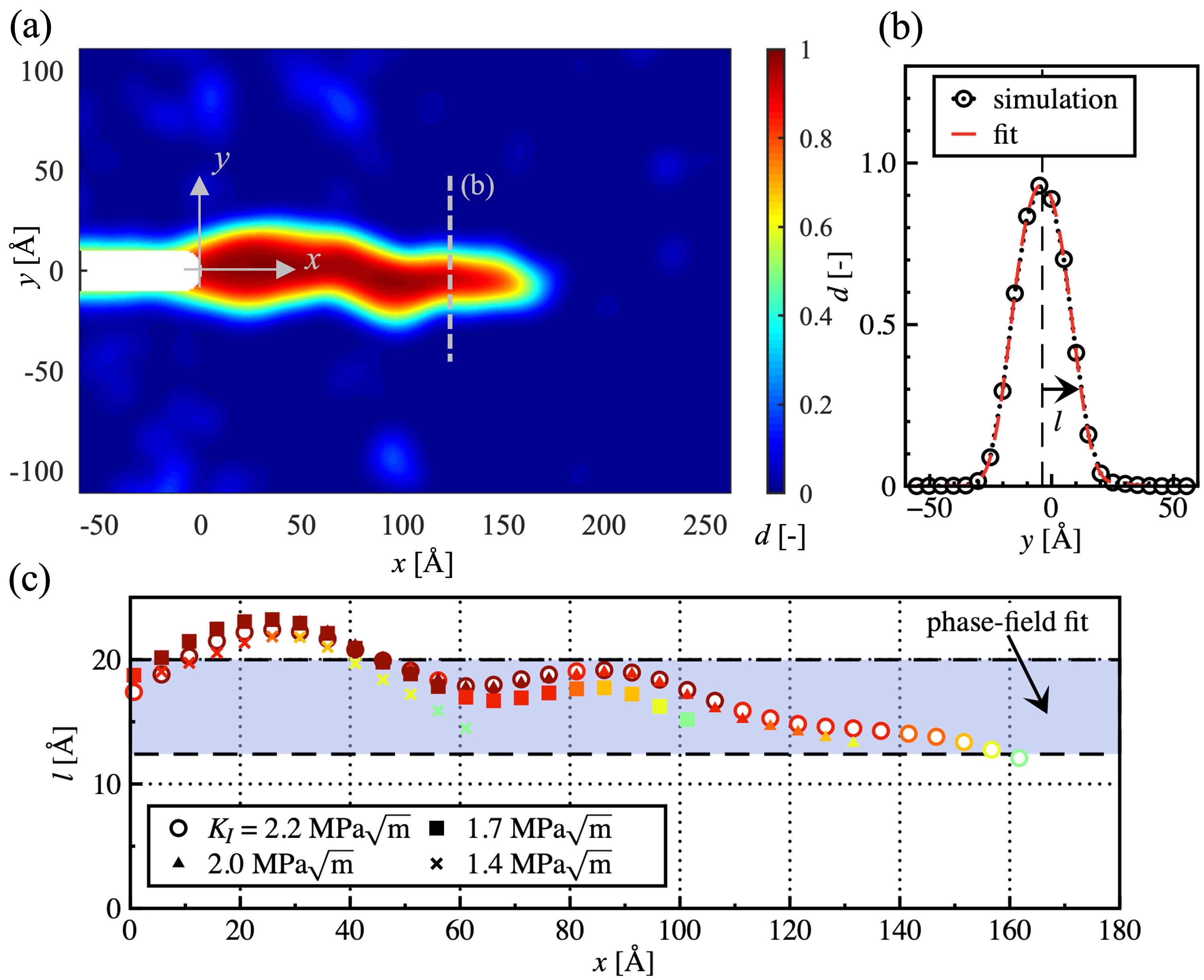}
	\caption{Damage field obtained from molecular simulations: (a) Distribution of damage in the middle plane of the sample. (b) Circles represent the damage profile along the $ y $ direction at $ x = 125$~\r{A}, fitted with a Gaussian function of maximum height $ d_{\rm{max}} $ and width $ l $. (c) Variation of the fitted width $ l $ along the crack and under different global loading states, color-coded by $ d_{\rm{max}} $ according to the colorbar. The blue region indicates $ l_c $ identified using the Finite Element Update (FEMU) scheme based on the phase-field formulation (see eq.~\ref{eq:PFSurf}).}
	\label{fig:damage}
\end{figure}

Fig.~\ref{fig:damage}a illustrates the damage distribution from molecular simulations in an $xy$ plane slice of the simulation box after fracture propagation over ca., 150~\r{A}. Fig.~\ref{fig:damage}b shows a cross section along $y$ at $x=125$~\r{A} with a Gaussian fit from which the damage width $l$ is derived. Finally, Fig.~\ref{fig:damage}c presents the damage width $l$ along the crack. The width varies but typically equals 20~\r{A} along the crack faces. The color map indicates maximum damage values along the $y$ axis. Of course, for consistency, the CG width $w$ must be kept small compared to the width of the damage zone $l$. 

As shown in detail in \ref{sec:leng}, we found that for $w = 3–5$~\r{A}, the mean value of $l$ remains the same, with only the fluctuation decreasing. At $w = 8$~\r{A}, the mean value starts to increase slightly, but this increase remains within the order of 10~\%. As $w$ increases beyond 8~\r{A}, the width of the damage zone, $l$, increases linearly with $w$. Subsequently we use the value $w = 8$~\r{A} which is approximately 3 times the size of a SiO$_4$ tetrahedron.                  

The coarse-grained atomic scale simulations demonstrate that damage around the crack is diffused rather than localized, extending into the material beyond the immediate crack tip. The diffusion width $l$ is notably larger than the CG width used to transition from atomic to continuum scales, indicating that the CG width is sufficient to capture local damage and small enough to expose the actual physical spread of damage. As shown Fig.~\ref{fig:damage}c, the damage diffusion length grows as the crack propagates but ultimately converges to a saturated value where the crack is fully open.
Specifically, for fully opened cracks, the damage zone varies in width between approximately 16 and 23 \r{A}, reflecting a non-uniform distribution of damage that suggests its extent may be influenced by structural heterogeneities such as the ring structure, which governs the brittleness and the differences between short-range and medium-range order in glasses \cite{yue2015}.  This local variability suggests that the diffusion length of damage is sufficiently small to be influenced by the underlying structural heterogeneity, and can therefore be regarded as a local material property rather than a global parameter.

Interestingly, the FEMU produced a damage width $l_c$ ranging from 12 to 20~\r{A}, depending on the loading state. This result correlates well with the observed coarse-grained damage width, even though $l_c$ is treated as a global parameter in the FEMU. This correlation suggests that, while damage diffusion is locally variable, the global parameter $l_c$ effectively captures the average behavior of the damage width. When solving the homogeneous phase-field solution \cite{Molnar2022EFM}, a comparable $\approx2$~nm regularization length was found based on experimental measurements.

\subsection{Local plasticity}

\begin{figure}
	\includegraphics{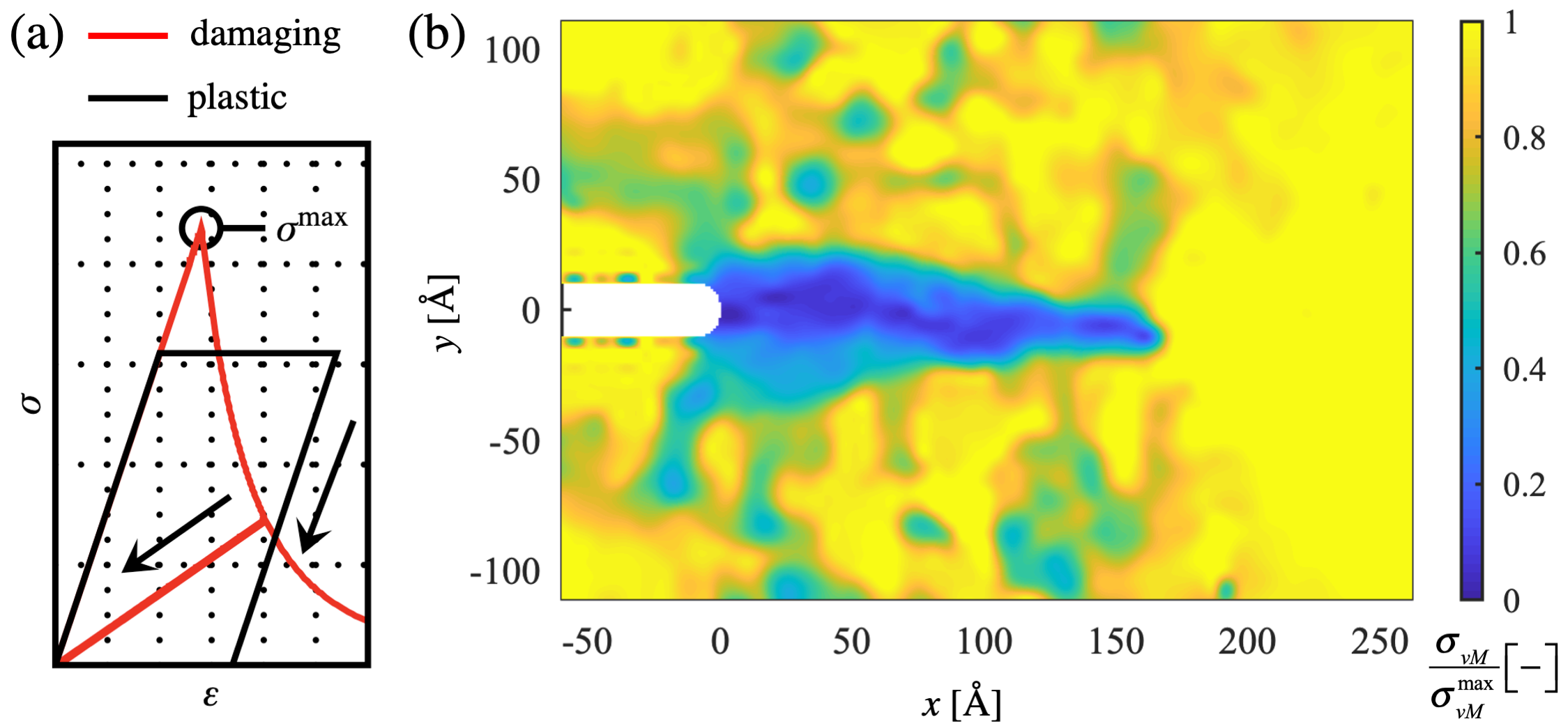}
	\caption{(a) Principal difference between damaging and plastic response. (b) Maximum stress ratio at $K_I=2.2$~MPa$\sqrt{m}$.}
	\label{fig:SvMmax}
\end{figure}

There are two essential phenomenological differences between damaging and ductile behavior: (i) in damage, no plateau is observed; instead, the stress reaches a maximum and then decreases to zero with further loading; (ii) the elastic stiffness in damage is lost, while in plasticity it is conserved upon unloading. These behaviors are illustrated in Fig.~\ref{fig:SvMmax}a.

Furthermore, to demonstrate that plastic effects are minimal during crack propagation, we present the von Mises stress divided by the maximum value throughout the entire loading process for a crack in Fig.~\ref{fig:SvMmax}b. It is evident that in the damaged zone, this ratio tends to zero, indicating damage. In contrast, if the material were ductile, the ratio would have remained around 1. Furthermore, this is reinforced by the energy balance presented in Section~\ref{sec:discus}, which shows that the unaccounted energies are minimal.

\subsection{Free surface energy}

The free surface energy ($2\gamma$) was determined using a smaller cubic sample with a side length of 100~\r{A}. We found that, the identified values to be independent of the sample size once it exceeded the 50~\r{A} box length.

We determined $2\gamma = 2.8 \pm 0.2$~J/m$^2$. This value is somewhat larger than room temperature experimental measurements including humidity ($\approx 0.5$ J/m$^2$) \cite{Wiederhorn69,Sarlat06,Kimura2015} but in agreement with calculated temperature variations~\cite{Yu2018} and correlates well with simulations using more sophisticated potentials \cite{Rimsza2018}.

Having access to local quantities gives us a unique opportunity to identify the origin and the distribution of the changes due to the newly formed surface. 

By looking at atomistic displacements, we observe that after cutting, equilibrium is reached through atomic displacements localized within a few interatomic distances from the surface. This finding aligns well with experimental observations of surface relaxation in crystals, where an exponential decay over the first few atomic layers has been reported \cite{Desjonqueres2012}. 

\begin{figure}
	\includegraphics{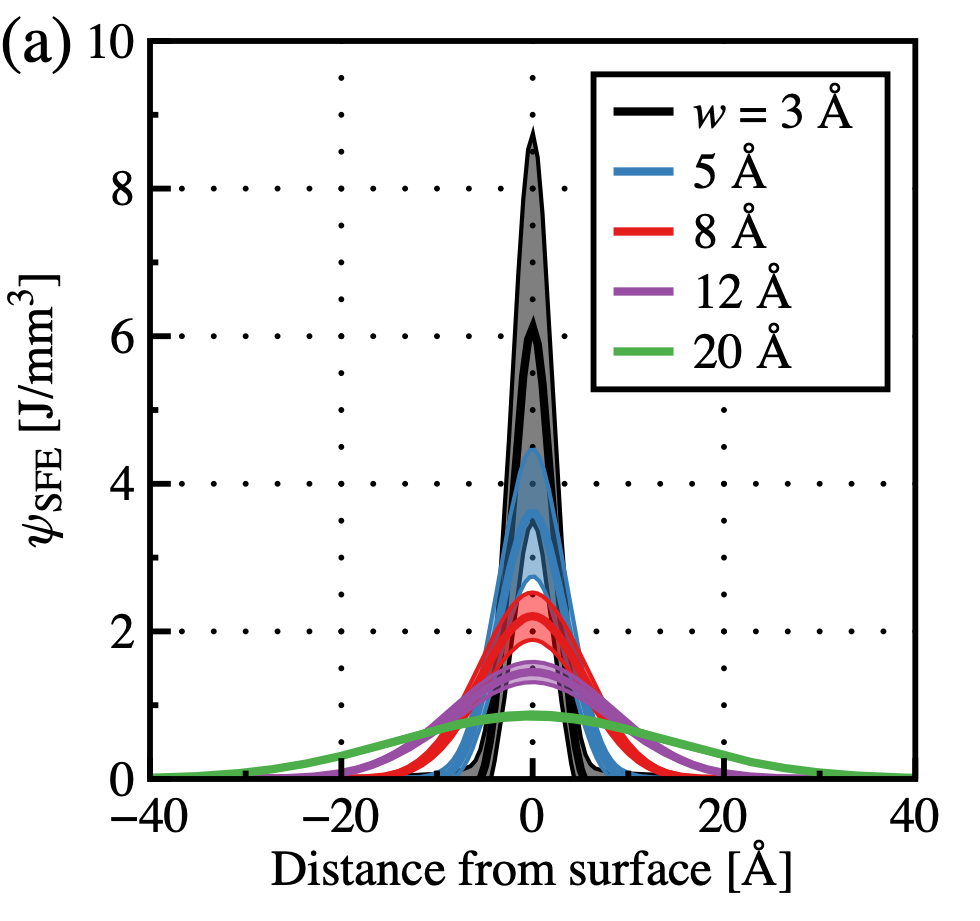}
	\includegraphics{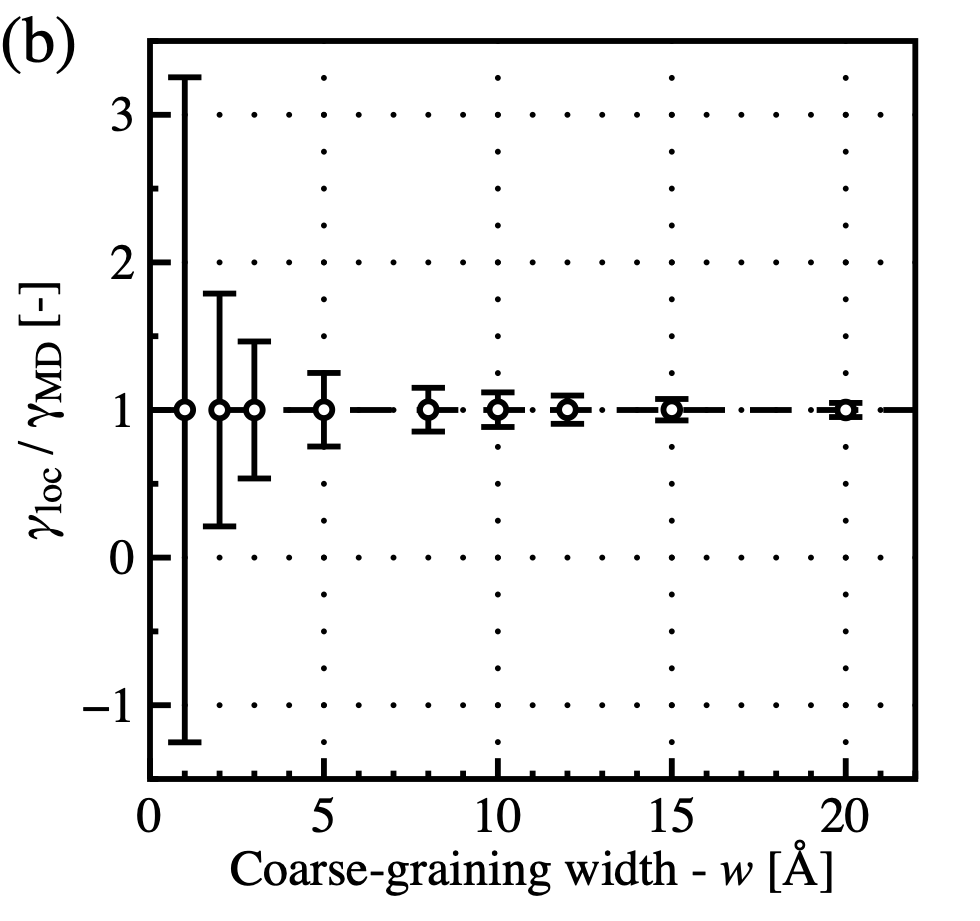}
	\caption{Distribution of free surface energy as a function of coarse-graining width. (a) Profiles of free surface energy along the $y$ axis for various coarse-graining widths. (b) Mean value and standard deviation of the local free surface energy $\gamma$, normalized by the global value.}
	\label{fig:cgw_FSE}
\end{figure}

Secondly, the local change in potential energy can be determined using a coarse-graining technique to estimate its spatial distribution. The coarse-graining was performed on the initial coordinates to maintain continuity, and the potential energy density values shown in Fig.~\ref{fig:cgw_FSE}a were shifted to the center for clearer visualization.

Fig.~\ref{fig:cgw_FSE}a shows that as a function of coarse-graining width, the profile of $d\psi_{\rm{SFE}}$ changes and reflects the coarse-graining function. This result indicates that the energy change is localized near the surface, and this change does not diffuse into the sample. However, when integrating this change along the $y$ axis, we recover a local $2\gamma$, whose mean value and standard deviation are displayed in Fig.~\ref{fig:cgw_FSE}b, normalized by the global value without coarse-graining. It is important to observe that the mean value, and thus the overall sum, is independent of the coarse-graining width; only the finer details are lost when $w$ is increased. Consequently, when determining the overall change in surface energy, the coarse-graining procedure does not affect the results.

This increase in energy is a clear signature of a newly created surface. Unfortunately, in a mechanically deformed sample, changes in potential energy are masked by elastic deformation, \ie the strain energy. Nevertheless, for purely elastic deformation, the potential energy $\psi_{\mathrm{pot}}$ is expected to be equal to the elastic energy $\psi_{\varepsilon}$ defined in eq.~\ref{eq:psi_eps}. In crack propagation, structural evolution—typically not accessible through purely elastic deformation—is reflected in deviations of $\psi_{\mathrm{pot}}$. We therefore introduce the difference, denoted

\begin{equation}
	\label{eq:FSE}
	\psi_{\rm{FSE}} = \psi_{\rm{pot}} -\psi_{\varepsilon}.   
\end{equation}

\noindent
reflecting the isolated free surface energy.


\begin{figure}[thb!]
	\includegraphics[width=0.75\textwidth]{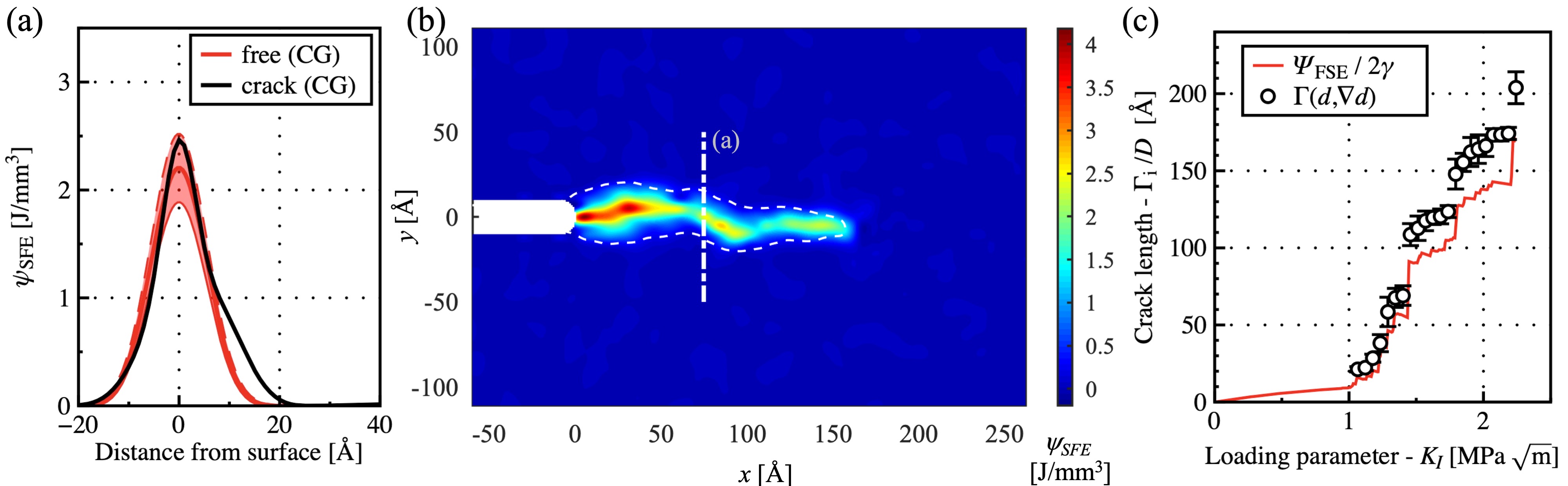}
	\caption{(a) Profiles of free surface energy for different models. In red, values for the dissected sample are shown with a CG width of 8~\r{A}. In black, results from the cracked and coarse-grained sample are shown at $x=75$~\r{A}. (b) Free surface energy distribution at $K_I=2.2$~MPa$\sqrt{m}$. (c) Crack surface as a function of loading derived from $\Psi_{\rm{FSE}}$ and from phase-field calculations.}
	\label{fig:surface}
\end{figure}

Fig.~\ref{fig:surface}a shows the distribution of $\psi_{\rm{FSE}}$ along the normal to the surface for the dissected sample. After coarse-graining, the profile width of $\psi_{\rm{SFE}}$ is observed to be 8~\r{A}, reflecting the convolution of the highly localized peak with the width $w$. A coarse-grained profile for the cracked samples, taken along the line indicated in (b), is also shown. The two coarse-grained $\psi_{\rm{FSE}}$ distributions are similar. This observation suggests that the $\psi_{\rm{FSE}}$ calculated in the fractured sample is the energy change due to the formation of the free surface. Fig.~\ref{fig:surface}b depicts the free surface energy during crack propagation.

Since the energy localization is consistent across different sample conditions, it appears that the free surface energy $2\gamma$ could also be an effective measure for assessing the area of newly formed cracks. Fig.~\ref{fig:surface}c displays crack length calculated from the global sample free surface energy $\Psi_{\rm{FSE}}$ divided by $2\gamma$. It is compared with the crack length determined by the FEMU based on the damage approach eq.~(\ref{eq:PFSurf}). This agreement between the two methods validates the use of $\psi_{\rm{FSE}}$ as a reliable metric for quantifying crack propagation, providing an alternative to more traditional methods and offering precise measurements that align well with continuum-based models.

\subsection{Structural signature}

Atomic-scale simulations are a powerful tool because they provide not only global responses but also allow the user to probe the local structure, offering a unique way to identify the structural signature underlying global phenomena. Nevertheless, these results should be treated as material-specific and should not be generalized without further validation. Rather, they should serve as a basis for additional analyses when investigating other materials.

In order to identify the structural rearrangements associated with crack opening, two local quantities were calculated. As expected during the opening of the network, changes were anticipated in both the ring structure of the glass and the coordination number of the network-forming silicon atoms.

\begin{figure}
	\includegraphics{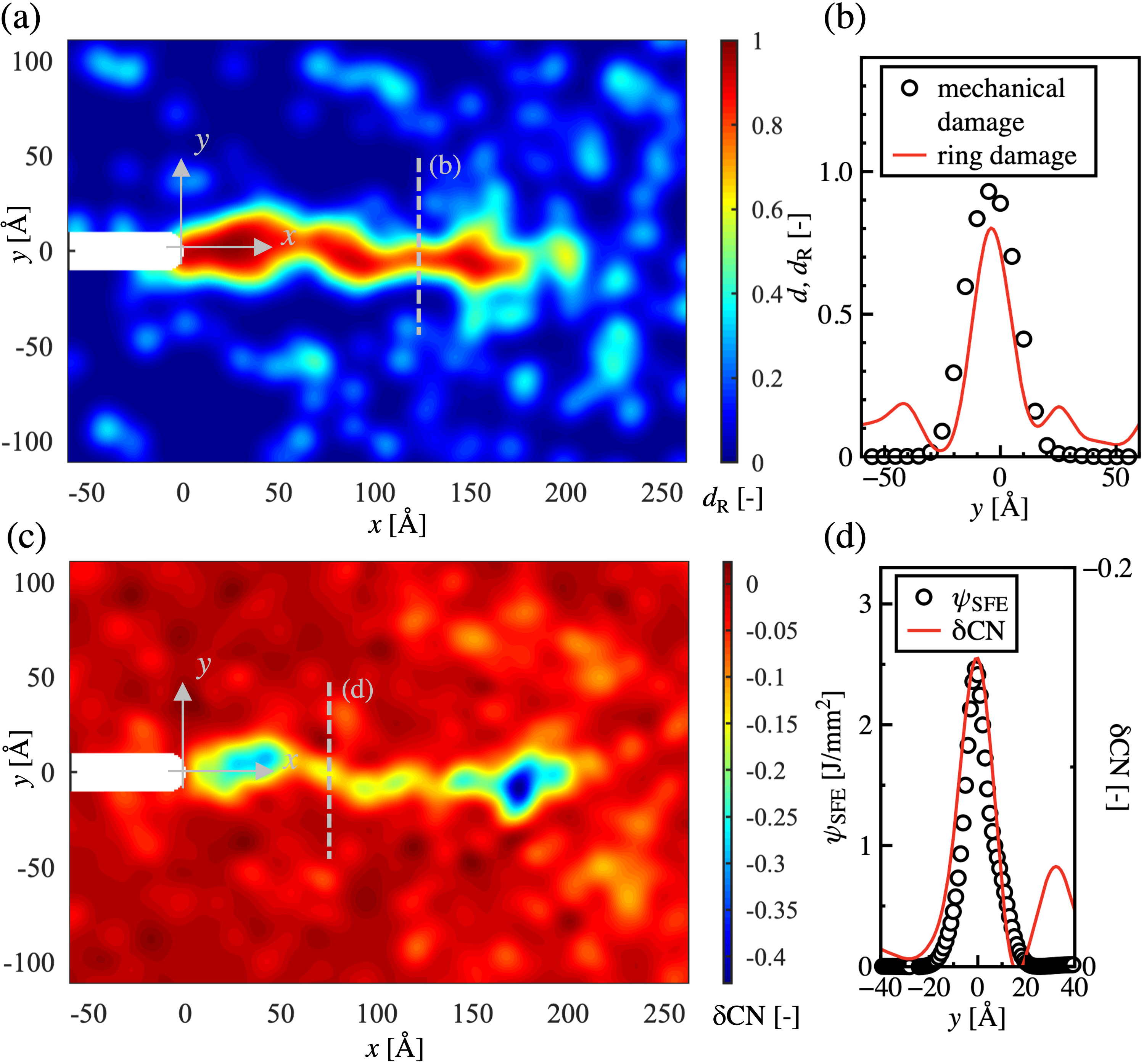}
	\caption{(a) Coarse-grained fraction of broken rings. (b) Vertical profiles of ring damage and phase-field damage at $x = 125$~\r{A}. (c) Coarse-grained coordination number change, computed from silicon atoms. (d) Vertical profiles of coordination number change and free surface energy at $x = 75$~\r{A}. All data correspond to a loading condition of $K_I = 2.2$~MPa$\sqrt{\mathrm{m}}$.}
	\label{fig:Str}
\end{figure}

To quantify changes in the ring structure, we employed the algorithm proposed by Le~Roux \& Jund \cite{LeRoux2010} to identify individual rings and monitor their evolution. Initially, each atom was assigned a value of $d_{\rm{R}} = 0$. If, compared to the initial configuration, a ring was no longer present in the deformed structure, all atoms belonging to that ring were assigned $d_{\rm{R}} = 1$. If an atom had already been associated with a broken ring, its value remained 1. Finally, a coarse-graining procedure was applied using a Gaussian smoothing kernel with a width of $w = 8$~\r{A}, resulting in Fig.~\ref{fig:Str}(a).

The change in coordination number, $\delta \mathrm{CN}$, was computed in a similar manner. First, the number of oxygen atoms within 1.7~\r{A} of each silicon atom was counted in both the initial and deformed configurations. These local coordination values were then coarse-grained using the same width $w = 8$~\r{A}. The difference between the initial and deformed coarse-grained values yielded the $\delta \mathrm{CN}$ field, shown in Fig.~\ref{fig:Str}(c).

When visualizing slices of each type of structural change, it becomes evident that ring damage exhibits a broader spatial extent, resembling phase-field damage. In contrast, changes in coordination number are more localized and narrow, akin to free surface energy distributions.

As a consequence, our analysis of the local structure revealed that diffuse damage is primarily associated with changes in the ring structure, whereas variations in free surface energy correlate with changes in the coordination number of silicon atoms.

\section{Energy equilibrium}
\label{sec:discus}

As noted in the introduction, experimental measurements have shown that the free surface energy resulting from surface relaxation accounts for only approximately 20\% of the energy required to open a crack. This section casts the free surface energy and the damage energy we have calculated at the atomic scale into a global energy balance to identify the origin of this discrepancy.

\begin{figure}
	\includegraphics{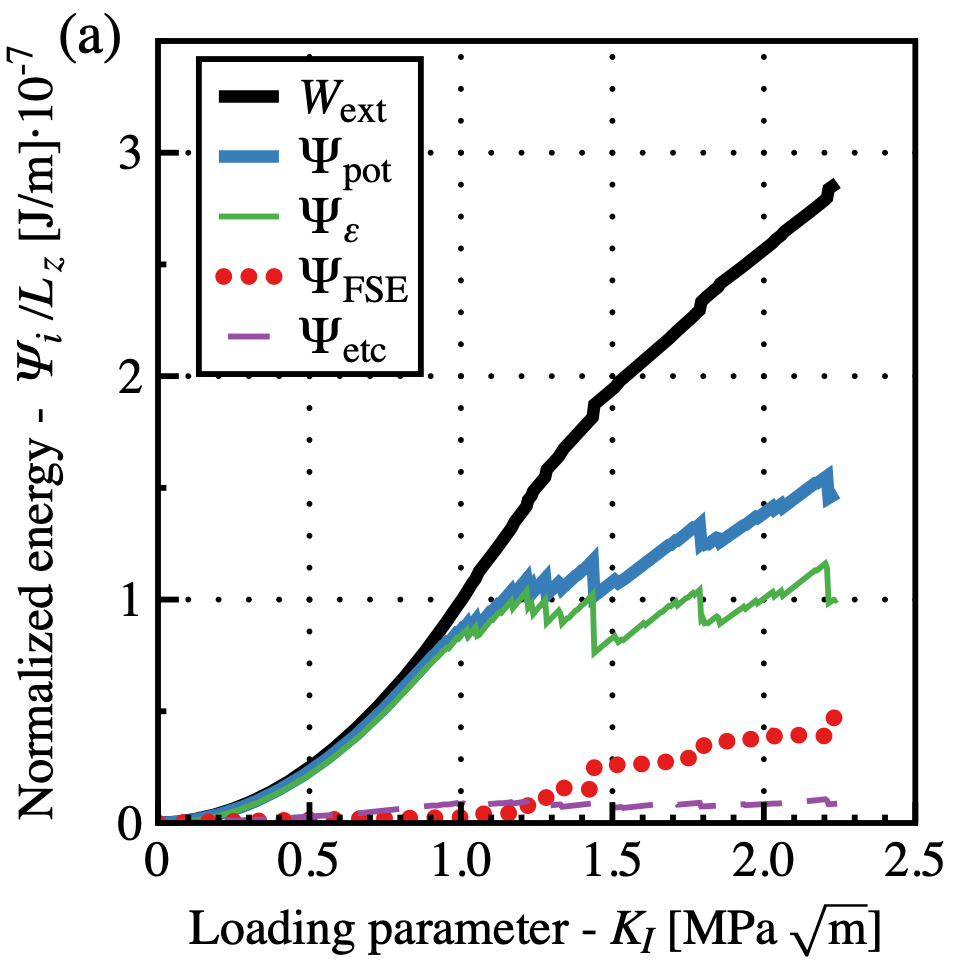}
	\includegraphics{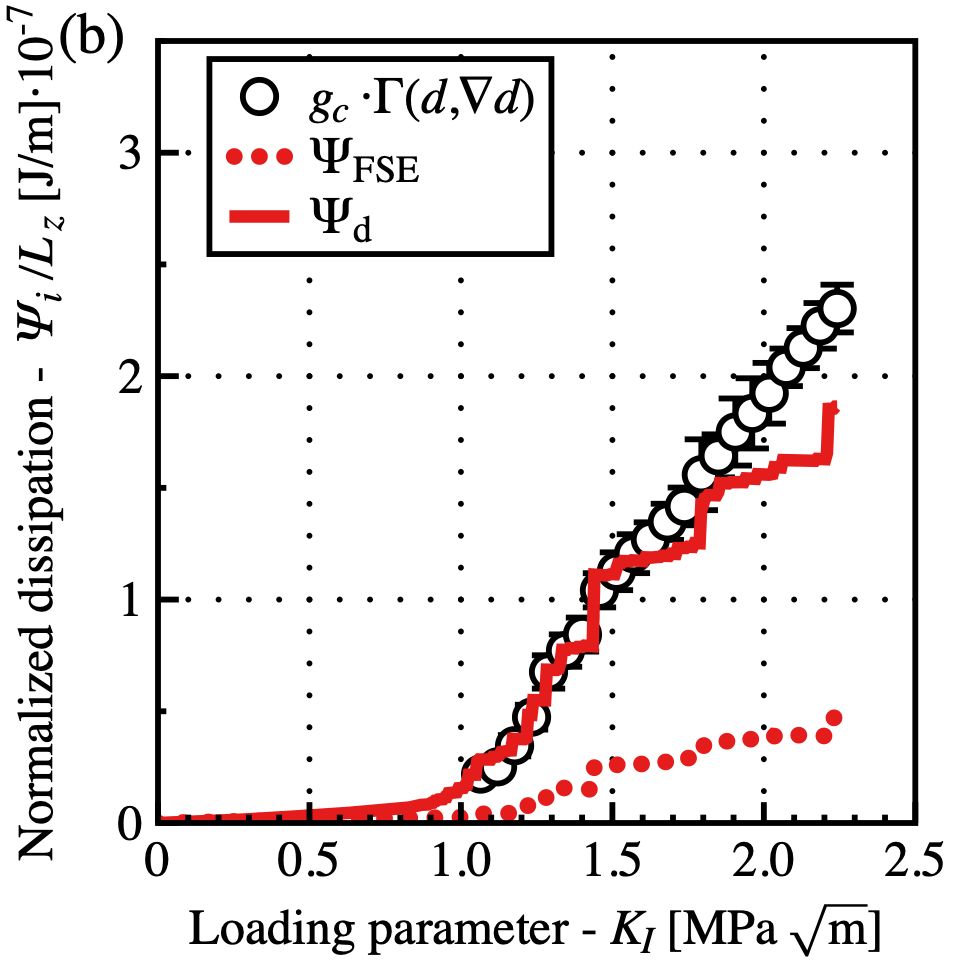}
	\includegraphics{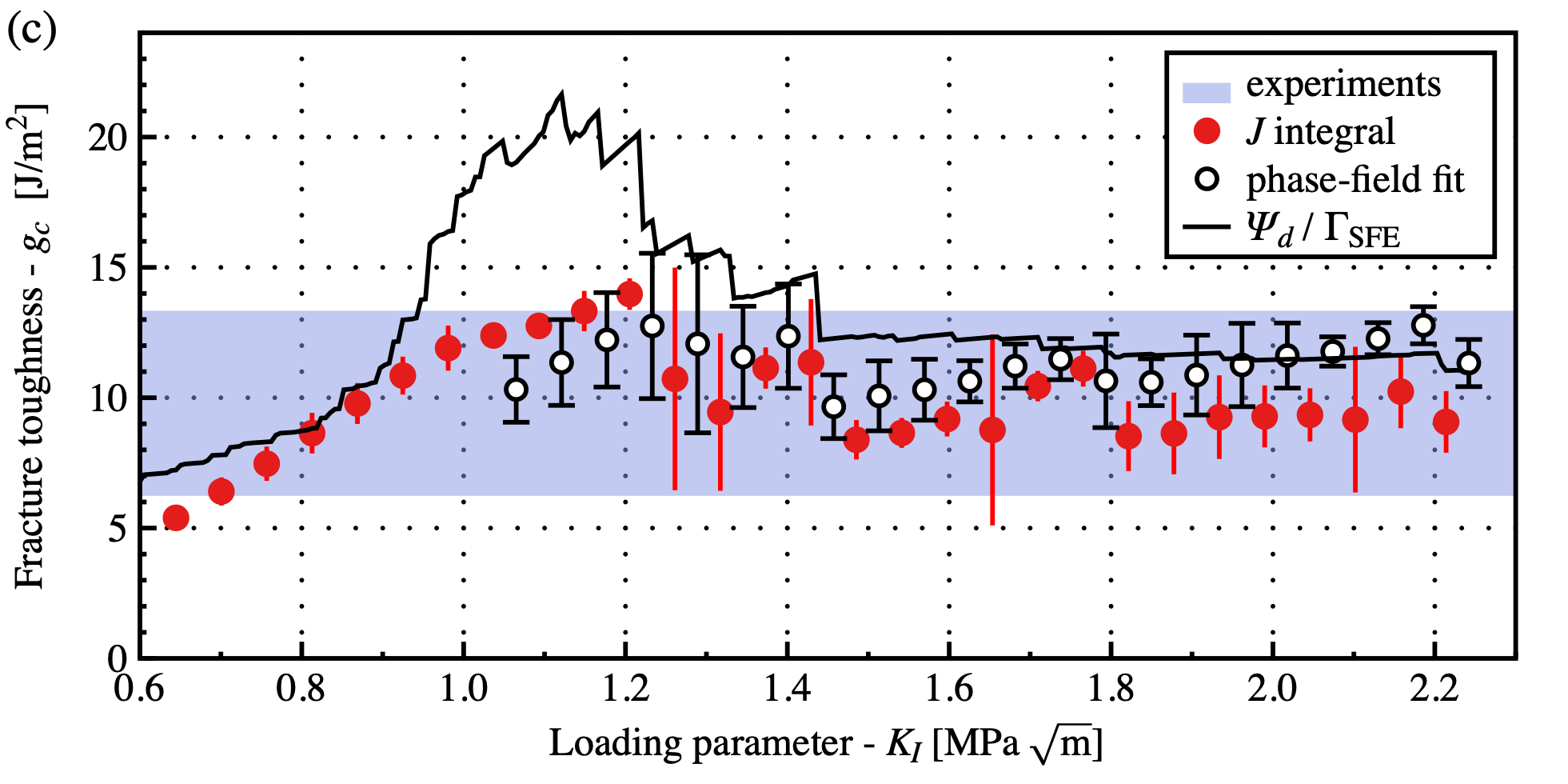}
	\caption{Energy balance of the cracked sample as a function of the loading parameter.
	We note that $K_I$ loses its precise physical significance as the crack tip advances and should be regarded solely as a loading parameter. 
	(a) Various energy contributions normalized by the thickness ($L_z$): external work ($W_{\mathrm{ext}}$), potential energy ($\Psi_{\mathrm{pot}}$), elastic strain energy ($\Psi_{\varepsilon}$), surface free energy ($\Psi_{\mathrm{SFE}}$), and other unquantified contributions (e.g., plasticity, folding, and nonlinear elasticity). 
	(b) Dissipation computed from the energy balance ($\Psi_d$) and crack surface energy obtained from phase-field calculations. 
	(c) Fracture toughness evaluated using different methods. The blue shaded region indicates the range of experimental measurements reported in Refs.~\cite{Wiederhorn69,Lucas1995,King2007,Mueller2015}.}
	\label{fig:energy}
\end{figure}

Fig.~\ref{fig:energy}a illustrates the various energy contributions as a function of loading. The external work, $W_{\rm{ext}}$, is partitioned into several components: the elastic strain energy, $\Psi_{\epsilon}$, and the free surface energy (FSE), $\Psi_{\rm{FSE}}$, whose sum is represented as $\Psi_{\rm{pot}}$. Additionally, there is a small non-linear contribution, $\Psi_{\rm{etc}}$, and a significant unquantified remainder.
The non-linear energy contribution is determined by isolating the portion of the elastic energy density that exceeds the local potential energy, which ideally remains positive under the assumption of linear elasticity. This excess energy contribution, is then subtracted from the total elastic energy. Notably, this contribution (represented by the dashed line) is small, suggesting that energy dissipation due to shear plasticity, densification or non-linear elasticity is indeed negligible. Under tensile loading followed by fracture, these mechanisms contribute only marginally to the overall energy dissipation within the material.

The difference between the external work and the elastic strain energy $W_{\rm{ext}} - \Psi_{\epsilon}$ can be interpreted as the total dissipation caused by fracture, denoted as $\Psi_{\rm{d}}$. This dissipated energy is shown in Fig.~\ref{fig:energy}b (solid red curve) and compared to the dissipation derived from the FEMU fit through eq.~\ref{eq:EngInt}. Although the two methods are very different (the FEMU does not incorporate explicit knowledge of the external work), they provide very consistent results, highlighting the robustness of the determination of the total energy dissipation. Also shown in the same graph is the FSE, which clearly constitutes only a portion of the total dissipation.

Finally, Fig.~\ref{fig:energy}c presents the fracture toughness ($g_c$) as a function of loading, calculated using three distinct methods: the FEMU phase-field fit (open circles), the energy balance approach ($\Psi_d / \Gamma_{\rm{SFE}}$) obtained from the coarse-grained atomic scale simulations and the J integral based on the coarse-grained stress and displacement fields (red filled circles). The J-integral is calculated as described in Section~\ref{sec:Jint}.
	
It is evident that the critical fracture surface energy remains consistent across the three calculation methods. The resulting value yields a ratio of $2\gamma/g_c \approx 23$\%: this value, which is determined at the atomic scale, matches the experimental findings \cite{Wiederhorn69,Lucas1995,King2007,Mueller2015}.

\section{Conclusion}
\label{sec:conclus}
This paper provides new insight into the origin of fracture toughness in amorphous silica, an archetype of brittle materials. We investigated the mechanisms responsible for the difference between free surface energy and fracture toughness in amorphous silica at the crossover between the atomic and the continuum
scales. Using a combination of atomic scale simulations and phase-field modeling, we quantified the different contributions to energy dissipation during fracture. From a comprehensive energy balance analysis, we were able to clearly isolate the free surface energy term and identify an additional contribution. Our analysis shows that this contribution arises almost exclusively from damage rather than plastic deformation, contradicting the conventional assumption~\cite{Wiederhorn69}. Notably, our calculations predict that this damage contribution exceeds the free surface energy by approximately a factor of 4, in agreement with experimental measurements.

These findings highlight that traditional views of fracture in brittle materials may overlook the subtle nature of mechanical dissipation in fracture of some amorphous materials. In particular they emphasize the dominant contribution of diffused damage in silicate glasses, aligning with experimental observations. These results also open new perspectives beyond silicate glasses. For example, it would be interesting to assess if these concepts apply to metallic glasses~\cite{lewandowski2005intrinsic} which are known to span the transition from brittle to ductile fracture. Understanding how these mechanisms couple to fatigue or to dynamic fracture is also a question worth investigating. The phase-field description is uniquely qualified to identify damage in the presence of plasticity. Therefore, it offers a more nuanced perspective for future materials design, particularly in optimizing the mechanical properties of brittle materials like amorphous silicates where, \emph{e.g.,} the addition of network modifiers may increase the contribution of plasticity.

\section*{Acknowledgment}
This research was funded, in part, by French Research National Agency program GaLAaD (ANR-20-CE08-0002).

\section*{References}
\bibliography{xampl}
\bibliographystyle{elsart-num}

\newpage

\appendix

\section{Effect of convolution length}
\label{sec:leng}

One of the main findings of this paper is the identification of an emerging length scale from molecular simulations using a Gaussian convolution. This homogenization process starts with a finite length, making it crucial to demonstrate that the results presented are essentially independent of the specific parameters used in the homogenization.

In this section, we explore how the width of the coarse-graining affects the results discussed in the main body of the article.

There are three key aspects that may be influenced by the homogenization process: (i) the elastic behavior, which indicates at what scale discrete atoms can be approximated as a continuum; (ii) the diffusion width of damage; and (iii) the effect of the free surface on local potential energy.

Our findings indicate that for a minimal length scale of  $w=8$~\r{A}, the elastic strain energy matches the local potential energy. This result aligns well with our previous findings on local elasticity. Additionally, we show that a width of $w=8$~\r{A} has minimal impact on damage diffusion. Lastly, we demonstrate that the free surface energy is highly localized on the surface, implying that while varying $w$ may alter the local maximum amplitude, it does not affect the overall sum of the energy change. Detailed explanations are provided below.

\begin{figure}
	\includegraphics{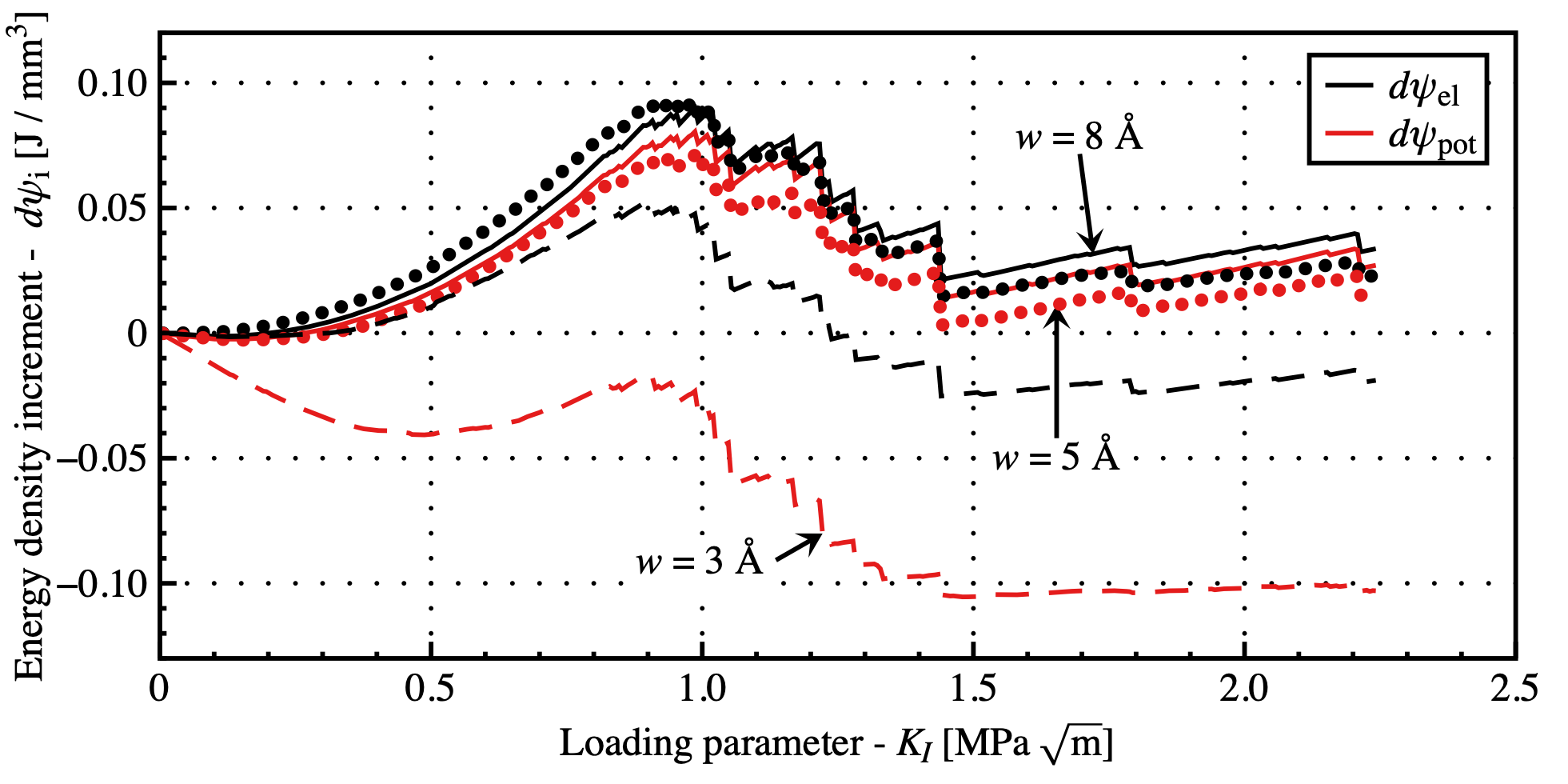}
	\caption{Impact of coarse-graining width on local elastic and potential energy densities in the material's elastic state.}
	\label{fig:cgw_engpot}
\end{figure}

Fig.~\ref{fig:cgw_engpot} depicts the local strain energy density ($d\psi_{\rm{el}}$) and potential energy density ($d\psi_{\rm{pot}}$) in black and red, respectively. The strain energy density is calculated from the stresses and strains, while the potential energy is coarse-grained from the atomistic values. These values are shown as a function of the global loading at a position in the sample that remains elastic during deformation. The descent of the curves indicates unloading, not damage nor plasticity. Under elastic conditions, these two values are expected to match. However, as illustrated in Fig.~\ref{fig:cgw_engpot}, if the homogenization length is too small, neither quantity is well-defined. When calculating the surface free energy ($d\psi_{\rm{SFE}} = d\psi_{\rm{pot}} - d\psi_{\rm{el}}$), this can result in a negative value, particularly in regions of the model where it should be zero, which is clearly not possible. Therefore, elastic analysis provides a lower bound for the coarse-graining width, which is in the range of $w = 6–8$~\r{A}. We note that a similar length scale, $w = 8$~\r{A}, was found where local elasticity begins to be well-defined~\cite{Molnar2016JNCS}. This was explained by micro-irreversible rearrangements present in amorphous materials even in the initial elastic stage, which are included in the slope of the experimentally measured stiffness.

\begin{figure}
	\includegraphics{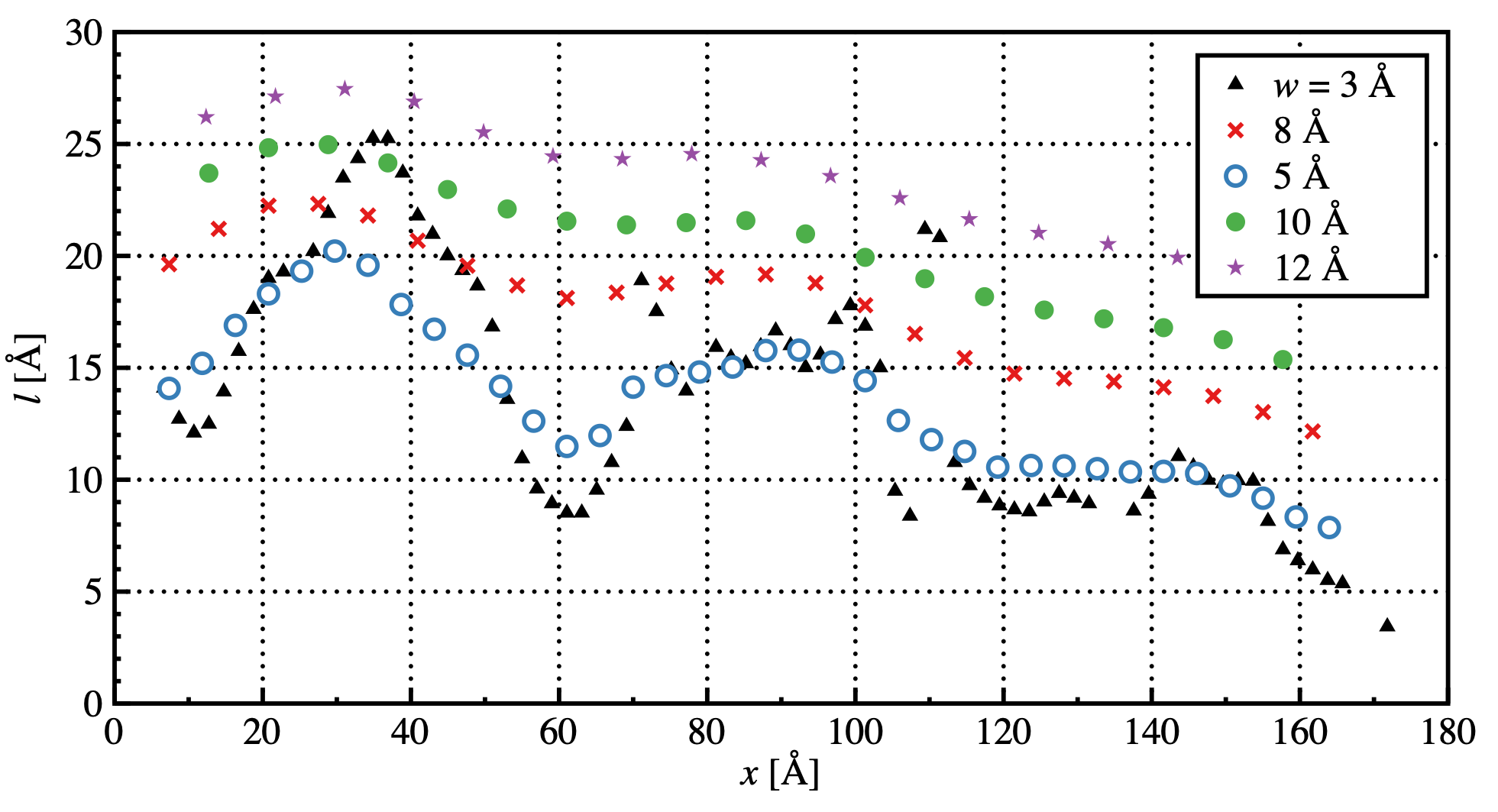}
	\caption{Effect of coarse-graining width on the damage diffusion width at $K_I=2.2$~MPa$\sqrt{m}$ in the middle plane.}
	\label{fig:cgw_damage}
\end{figure}

While for elasticity, the larger the coarse-graining width, the better the results represent a continuum, for localized damage, it hinders the identification of the extent of the diffused damage. Fig.~\ref{fig:cgw_damage} presents the width of the damage zone in the middle plane for $K_I = 2.2$~MPa$\sqrt{m}$ with various coarse-graining widths along the crack.

It is clearly visible that for $w = 3–5$~\r{A}, the mean value remains the same, with only the fluctuation decreasing. At $w = 8$~\r{A}, the mean value starts to increase slightly, but this increase remains within the order of 10~\%. Beyond this, as $w$ increases, the width of the damage zone, $l$, increases linearly.

To keep negative surface energies minimal and stresses well-defined, we accepted a slight effect on the damage diffusion and chose $w = 8$~\r{A} for subsequent analysis. It is important to note that the exact value of $l$ might be slightly smaller. This consideration is crucial as the exact value depends on the degradation function and the phase-field description.

\section{J integral}
\label{sec:Jint}

The J integral, a contour integral, was originally proposed to deduce the energy liberated from an elastic body upon the potential advancement of a sharp crack within a homogeneous and isotropic domain. Essentially, the contour integral provides the difference between the work done on the contour and the energy stored in the solid. The resulting difference, assuming a unitary, straight, horizontal crack advancement, gives the energy release rate. Its advantage at the time was that it captured the additional energy dissipated by ductile deformation and attributed it to the increment of fracture toughness. Therefore, the method is applicable in ductile cases.

In this paper, we use the J integral to gain an approximate description and to verify the critical fracture toughness identified by the phase-field damage model.

\begin{figure}
	\includegraphics[width=0.23\textwidth]{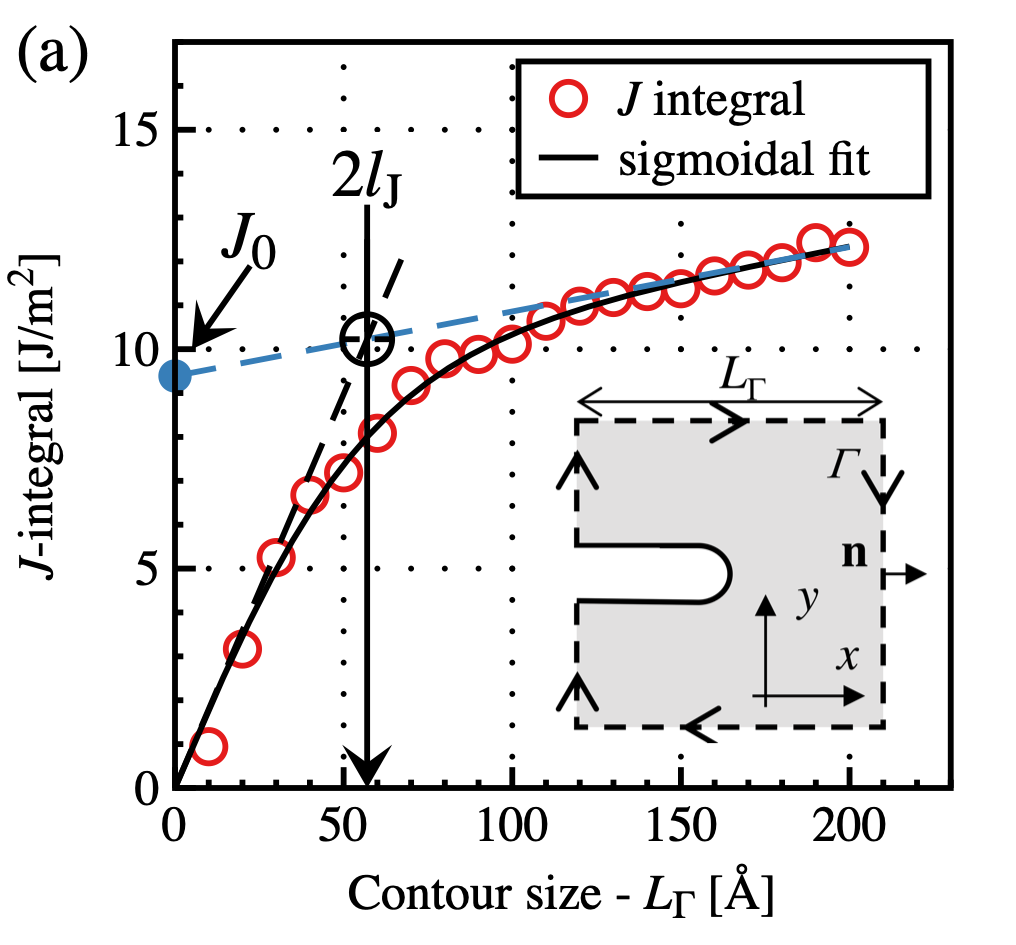}
	\includegraphics[width=0.23\textwidth]{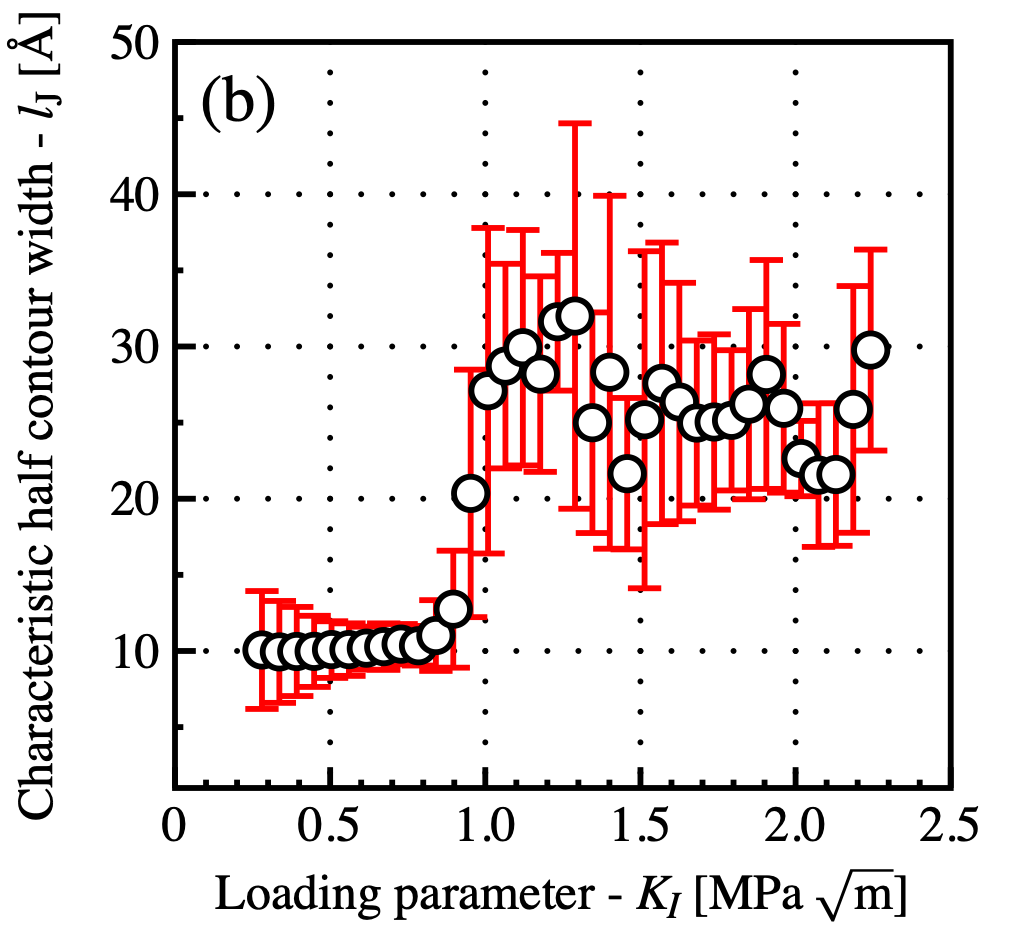}
	\caption{(a) Schematic illustration to calculate the J integral. (b) Characteristic half contour width as a function of loading.}
	\label{fig:J_integral}
\end{figure}

The contour integral is defined as:

\begin{equation}
	\label{eq:J}
	J = \int\limits_{{\Gamma _J}} {\left( {{\psi _{el}}{n_x} - {\bf{t}}\frac{{\partial {\bf{u}}}}{{\partial x}}} \right)} d\Gamma
\end{equation}

where ${\bf{t}} = {\bf{P n}}$ with $\bf{P}$ being the coarse-grained first Piola-Kirchhoff stress tensor and $\bf{n}$ the normal vector to the contour.

The contour integral was centered at the local maximum of the first principal stress peak in front of the crack. To approximate the effect of crack undulation multiple contours were used, with widths denoted as $L_\Gamma$. Finally, a sigmoidal fit, which transformed into a linear function, was applied:

\begin{equation}
	\label{eq:Jfit}
	J\left( {{L_\Gamma }} \right) = \frac{{{J_\infty }}}{{1 + {e^{ - b{L_\Gamma }}}}} + {L_\Gamma }c - \frac{{{J_\infty }}}{2}.
\end{equation}

This fitting procedure is displayed in Fig.~\ref{fig:J_integral}a.

Due to local damage and the inhomogeneous nature of the material, this contour integral is not well-defined at the crack tip. Additionally, the function exhibits a slight slope at larger contour sizes because the expected K-field is influenced by the model's boundary conditions. Therefore, the results shown in Section~\ref{sec:discus} display $J_0$, which is a linear interpolation of $J$ at $L_\Gamma = 0$. Interestingly, this fit allows us to identify the length where the integral is valid and the response behaves closely to linear elastic fracture mechanics. This is done by intersecting the initial slope and the line from the final slope of eq.~(\ref{eq:Jfit}), as shown in Fig.~\ref{fig:J_integral}a. The length $l_J$ is displayed in Fig.~\ref{fig:J_integral}b. It shows a similar size to where the material becomes treatable with linear elastic fracture mechanics when the contour becomes larger than the size of the diffused damaged zone.

\end{document}